\documentclass[twoside,english,american]{article}
\usepackage{spconf}
\usepackage[T1]{fontenc}
\usepackage[latin9]{inputenc}
\usepackage{float}
\usepackage{bm}
\usepackage{amsmath}
\usepackage{amssymb}
\usepackage{graphicx}

\makeatletter

\floatstyle{ruled}
\newfloat{algorithm}{tbp}{loa}
\providecommand{\algorithmname}{Algorithm}
\floatname{algorithm}{\protect\algorithmname}

\usepackage{float}
\usepackage{mathtools}
\usepackage{bm}
\usepackage{amsthm}
\usepackage{amsmath}
\usepackage{amssymb}
\usepackage{graphicx}
\usepackage{amsmath,epsfig}
\usepackage{cite}
\usepackage{dblfloatfix}
\usepackage{graphicx, amsmath}
\usepackage{subfigure}
\usepackage{cite}
\usepackage{algorithm}
\usepackage{algorithmic}
\newcommand{\re}{\text{red}}

\renewcommand{\fnum@figure}{Fig.~\thefigure}
\allowdisplaybreaks
\theoremstyle{plain}

\theoremstyle{definition}

\theoremstyle{remark}

\floatstyle{ruled}
\newfloat{algorithm}{tbp}{loa}
\providecommand{\algorithmname}{Algorithm}
\floatname{algorithm}{\protect\algorithmname}
\usepackage{babel}
\addto\captionsamerican{\renewcommand{\definitionname}{Definition}}
\addto\captionsamerican{\renewcommand{\remarkname}{Remark}}
\addto\captionsamerican{\renewcommand{\theoremname}{Theorem}}
\addto\captionsenglish{\renewcommand{\algorithmname}{Algorithm}}
\addto\captionsenglish{\renewcommand{\definitionname}{Definition}}
\addto\captionsenglish{\renewcommand{\remarkname}{Remark}}
\addto\captionsenglish{\renewcommand{\theoremname}{Theorem}}
\providecommand{\definitionname}{Definition}
\providecommand{\remarkname}{Remark}
\providecommand{\theoremname}{Theorem}
\title{Globally Optimal Beamforming Design for Downlink CoMP transmission with Limited Backhaul Capacity}

\name{Kien-Giang Nguyen$^{\ast}$, Quang-Doanh Vu$^{\ast}$, Markku Juntti$^{\ast}$, and Le-Nam Tran$^{\dag}$ \thanks{This work was supported in part by the Academy of Finland under projects Message and CSI Sharing for Cellular Interference Management with Backhaul Constraints (MESIC) belonging to the WiFIUS program with NSF, and Wireless Connectivity for Internet of Everything (WiConIE), and the HPY Research Foundation. This project has been co-funded by the Irish Government and the European Union under Ireland's EU Structural and Investment Funds Programmes 2014-2020 through the SFI Research Centres Programme under Grant 13/RC/2077.}}
\address{\small $^{\ast}$ Centre for Wireless Communications, University of Oulu, P.O.Box 4500, FI-90014, Finland; \\
\small Email: \{giang.nguyen, doanh.vu, markku.juntti\}@oulu.fi. \\
\small $^{\dag}$ Department of Electronic Engineering, Maynooth University, Ireland; Email: lenam.tran@nuim.ie.}
\ninept

\begin{document}
\maketitle
\ninept
\begin{abstract}
This paper considers a multicell downlink channel in
which multiple base stations (BSs) cooperatively serve users by jointly
precoding shared data transported from a central processor over limited-capacity
backhaul links. We jointly design the beamformers and BS-user link
selection so as to maximize the sum rate subject to user-specific
signal-to-interference-noise (SINR) requirements, per-BS backhaul
capacity and per-BS power constraints. As existing solutions for the
considered problem are suboptimal and their optimality remains unknown
due to the lack of globally optimal solutions, we characterized this
gap by proposing a globally optimal algorithm for the problem of
interest. Specifically, the proposed method is customized from a generic
framework of a branch and bound algorithm applied to discrete monotonic
optimization. We show that the proposed algorithm converges after
a finite number of iterations, and can serve as a benchmark for existing
suboptimal solutions and those that will be developed for similar
contexts in the future. In this regard, we numerically compare the
proposed optimal solution to a current state-of-the-art, which show
that this suboptimal method only attains 70\% to 90\% of the optimal
performance.
\end{abstract}
\begin{keywords}
Multicell cooperation, limited backhaul, sum rate maximization, discrete monotonic optimization.
\end{keywords}

\section{Introduction}

Due to the explosive growth of wireless devices and data services,
increasing network capacity has become a critical target for the current
and future wireless networks. As well concluded from pioneer research,
intercell-inteference is a key factor limiting the overall spectral
efficiency offered by wireless technologies \cite{Multicell-MIMO-Cooperative-Networks}.
Schemes turning interference from nuisance to the users' benefit,
such as cooperative multipoint (CoMP) processing has been proposed
\cite{marsch2011coordinated}. The central idea is to allow for joint
processing by multiple transmitters, thereby turning the interference
to useful signals \cite{Multicell-MIMO-Cooperative-Networks,marsch2011coordinated,HowdesignCOMP}.
 The two main forms of CoMP processing are interference coordination
and data sharing. In the latter, data for a specific is transmitted
from multiple base stations (BSs), which is the focus of this paper.

The promise of multicell BS cooperation is often seen under the assumption
that the backhaul network is able to deliver an enormous signaling
overhead from a central processor (CP) to all BSs.  In practice,
the capacity of the backhaul communications is limited, especially,
with wireless backhauling \cite{Survey_backhaul_tech}. Therefore,
a number of recent research efforts have investigated solutions to
optimize various performance measures in finite-capacity backhaul
networks for different system models. Those include, e.g., sum rate
maximization \cite{Dai2014}, joint backhaul and power minimization
\cite{FuxinKLau:2014:SP:BackhaulLimitedSDPRelax}, energy efficiency
maximization \cite{DerrickKwanNg:2012:JWCOM:EE_OFDM}, and backhaul
usage minimization \cite{JZhaoTonyQ:2013:JWCOM:CoMPBackhaul}. To
deal with the backhaul constraint in particular, the aforementioned
works have adopted a BS-user link selection scheme where each BS in
the network transmits data to only a proper subset of users in order
to reduce the backhaul consumed.

This paper is to explore the optimal performance of maximizing the
sum rate of cooperative multicell downlink under limited backhaul
capacity constraints. In particular, we propose a beamforming design
to maximize the achievable sum rate while satisfying per-BS backhaul
capacity, per-BS power constraints, and user-specific signal-to-interference-noise
(SINR) requirements. The latter condition ensures the minimum quality
of service for each user regardless of the user's location in a cell.
To cope with the backhaul limitation, a BS-user link selection scheme
is necessary and will be studied in this paper. Thus, the resulting
problem is in fact a joint beamforming and BS-user link selection
design which naturally leads to a mixed-Boolean non-convex program
(MBNP), which is generally known to be NP-hard and optimal solutions
are hard to derive. A suboptimal solution based on reweighted $l_{1}$-norm
was proposed for a similar problem in the considered context \cite{Dai2014}.
Although this approach yields solutions with reasonably low complexity,
its achieved performance in relation to the optimal one has not been
investigated yet. To fill this gap, we propose an algorithm that
achieves a global optimum of the design problem. The proposed method
is based on a discrete monotonic optimization (DMO) framework which
is difficult to connect to the original problem formulation. In particular,
the hidden monotonicity of the considered problem is exposed by a
novel proposed transformation. This then facilitates an efficient
customization of a discrete brach-reduce-and-bound (DBRB) method to
find a global optimum. Numerical results are provided to demonstrate
the convergence of the proposed optimal algorithm and the performance
gains over known methods.\vspace{-2mm}

\section{System Model and Problem Formulation}

\vspace{-1mm}Consider a multiple-input single-output (MISO) downlink
transmission of wireless systems where there are $B$ multiple-antenna
BSs, each is equipped with $M$ antennas, jointly serving $K$ single-antenna
users. Suppose that all BSs in the network are connected to a CP through
backhaul links of limited capacity. We also assume that the CP has
the data of all users and perfectly knows the associated channel state
information (CSI) and the necessary control information. The data
symbol for user $k$ is denoted by $d_{k}$, assumed to have unit
energy, i.e., $\mathbb{E}[|d_{k}|^{2}]=1$. Herein, we adopt linear
precoding, i.e., $d_{k}$ is multiplied with beamformer ${\bf w}_{b,k}\in\mathbb{C}^{M\times1}$
before being transmitted by BS $b$. Accordingly, under flat fading
channels, the received signal ${\bf y}_{k}\in\mathbb{C}^{M\times1}$
of user $k$ can be written as
\begin{equation}
{\bf y}_{k}={\textstyle \sum_{b=1}^{B}}{\bf h}_{b,k}{\bf w}_{b,k}d_{k}+{\textstyle \sum_{b=1}^{B}\sum_{j\neq k}^{K}}\mathbf{h}_{b,k}\mathbf{w}_{b,j}d_{j}+\sigma_{k},
\end{equation}
where ${\bf h}_{b,k}\in\mathbb{C}^{1\times M}$ is the (row) vector
representing the channel of $(b,k)$, and $\sigma_{k}\sim\mathcal{CN}(0,N_{0})$
is the additive white Gaussian noise at user $k$. For notational
simplicity, let ${\bf h}_{k}\triangleq[{\bf h}_{1,k},{\bf h}_{2,k},\ldots,{\bf h}_{B,k}]\in\mathbb{C}^{1\times MB}$
and ${\bf w}_{k}\triangleq[{\bf w}_{1,k}^{T},{\bf w}_{2,k}^{T},\ldots,{\bf w}_{B,k}^{T}]^{T}$
$\in\mathbb{C}^{MB\times1}$ be the aggregate vectors of all channels
and beamformers from all BSs to user $k$, respectively. We also denote
by ${\bf w}$ the beamforming vector encompassing all ${\bf w}_{k}$.
We further assume that single-user decoding is performed. In this
regard, the intercell-interference is treated as Gaussian noise, and
thus the SINR at user $k$ can be written as
\begin{equation}
\gamma_{k}({\bf w})\triangleq|{\bf h}_{k}{\bf w}_{k}|^{2}({\textstyle \sum_{j=1,j\neq k}^{K}}|{\bf h}_{k}{\bf w}_{j}|^{2}+WN_{0})^{-1}.
\end{equation}
The data rate of user $k$ is given by $R_{k}({\bf w})\triangleq W\log(1+\gamma_{k}({\bf w}))$,
where $W$ is the bandwidth. To simplify the notation, we will drop
$W$ in the sequel.

Let us denote by $x_{b,k}\in\{0,1\}$ the selection preference
variable where $x_{b,k}=1$ indicates that transmission link between
BS $b$ and user $k$ is active and $x_{b,k}=0$ otherwise. The backhaul
usage of BS $b$ is the sum of data streams of its served users given
by $C_{b}^{\text{BH}}\triangleq{\textstyle \sum_{k=1}^{K}}x_{b,k}R_{k}({\bf w})$
which is upper bounded by a link capacity $\bar{C}$, i.e., $C_{b}^{BH}\leq\bar{C}$.
To reduce the backhaul usage, BS $b$ can turn off transmissions to
some users. That is, the CP does not deliver those users' information
to BS $b$. Also, beamformers associated to inactive transmissions
are forced to be zero, i.e., ${\bf w}_{b,k}=0$ if $x_{b,k}=0$. To
capture this relation, we introduce the constraint $\|{\bf w}_{b,k}\|\leq x_{b,k}u_{b,k}$
where $u_{b,k}$ represents a soft power level of ${\bf w}_{b,k}$
and will be optimized under the considered power constraint.

Based on the above discussions, the problem of joint beamforming design
and BS-user link selection which maximizes the sum rate subject to
user-specific SINR requirements, per-BS backhaul capacity and power
constraints can be formulated as \begin{subequations}\label{Prob:Gen:Problem}
\begin{align}
\underset{{\bf w},{\bf x},{\bf u}}{\text{maximize}} & \ \ \ {\textstyle \sum_{k=1}^{K}}R_{k}({\bf w})\label{eq:obj:gen:prob}\\
\text{subject to} & \ \ \ {\textstyle \sum_{k=1}^{K}}x_{b,k}R_{k}({\bf w})\leq\bar{C},\ \forall b,k\label{eq:BH}\\
 & \ \ \ \gamma_{k}({\bf w})\geq\bar{\gamma}_{0},\forall k\label{eq:QoS}\\
 & \ \ \ \|{\bf w}_{b,k}\|_{2}^{2}\leq x_{b,k}u_{b,k},\ \forall b,k\label{eq:soft power}\\
 & \ \ \ {\textstyle \sum_{k=1}^{K}}u_{b,k}\leq\bar{P},\ \forall b,k\label{eq:power}\\
 & \ \ \ {\textstyle \sum_{b=1}^{B}}x_{b,k}\geq1,\ \forall b,k\label{eq:min:connectivity}\\
 & \ \ \ x_{b,k}\in\{0,\ 1\},\ \forall b,k,\label{eq:boolean}
\end{align}
\end{subequations}where $\bar{\gamma}_{0}$ is the per-user targeted
SINR, and $\bar{P}$ is the per-BS transmit power budget. We also
denoted ${\bf x}\triangleq[x_{1,k},\ldots x_{b,k},\ldots,x_{B,K}]^{T}\in\{0,1\}^{BK}$
and ${\bf u}\triangleq[u_{1,k},\ldots u_{b,k},\ldots,u_{B,K}]^{T}\in\mathbb{R}_{+}^{BK}$
. Herein \eqref{eq:min:connectivity} is added to ensure each user
is always served by at least one BS. Note that user-specific SINR
constraint \eqref{eq:QoS} can be represented as an SOC constraint
following the results in \cite{WieselEldar:06:LinearPrecoding} as
\begin{equation}
\mathbf{h}_{k}\mathbf{w}_{k}\geq\sqrt{\bar{\gamma}_{0}({\textstyle \sum_{j\neq k}^{K}}|\mathbf{h}_{k}\mathbf{w}_{j}|^{2}+WN_{0})}\ ,\ \Im(\mathbf{h}_{k}\mathbf{w}_{k})=0.\label{eq:QoS:SoCP}
\end{equation}
It is now clear that problem \eqref{Prob:Gen:Problem} belongs to
the class of MBNP due to the Boolean variable ${\bf x}$, the nonconvex
objective \eqref{eq:obj:gen:prob} and the nonconvex constraint \eqref{eq:BH}.
Recall that a suboptimal solution for this problem was proposed in
\cite{Dai2014} using a sparsity inducing norm approach. Thus, there
is a need to understand its achieved performance, and to see if there
is a room for improvement.\vspace{-2mm}

\section{Proposed Optimal Solution}

\vspace{-1mm}We derive an algorithm which globally solves \eqref{Prob:Gen:Problem}
by customizing a state-of-the-art global discrete optimization technique
namely discrete monotonic optimization. This framework has been applied
to find a global optimum of different MBNP problems in wireless communications
\cite{Monotonic-LinearDecoding,zhang2013monotonic,CheTuan-JointCoopBeamRelay}.
It is worth noting that although continuous monotonic optimization
(MO) \cite{tuy2005monotonic} can also handle discrete constraints
in \eqref{eq:boolean} by writing $x_{bk}\in\{0,1\}\Leftrightarrow x_{b,k}^{2}-x_{bk}\geq0,\ x_{b,k}\in[0,1]$.
However it potentially returns only approximate solutions by a finite
number of iterations. Thus, DMO has been developed in \cite{tuy2006discrete}
to compute exact optimal solutions. In this section, we will adapt
the discrete branch-reduce-and-bound strategy in DMO, to solve the
design problem \eqref{Prob:Gen:Problem}. Before proceeding further,
we remark that basic concepts of the MO such as \emph{increasing function},
\emph{normal cone} and \emph{box} are used throughout the rest of
this section. Their definitions can be found in \cite{tuy2005monotonic}
and are omitted in this paper due to space limitation.

The current formulation of the design problem is not amendable for
a direct application of the DBRB since \eqref{Prob:Gen:Problem} does
not hold the monotonicity property w.r.t. the involved variables.
Thus, a further proper translation is required. For this purpose,
let us introduce new slack variable $z_{k}$ and rewrite \eqref{Prob:Gen:Problem}
as \begin{subequations}\label{Prob:SE:epi}\vspace{-1mm}
\begin{align}
\underset{{\bf w},{\bf x},{\bf u},{\bf z}}{\text{maximize}} & \ \ \ {\textstyle \sum_{k=1}^{K}}z_{k}\\
\text{subject to} & \ \ \mathbf{h}_{k}\mathbf{w}_{k}\geq\sqrt{(e^{z_{k}}-1)({\textstyle \sum_{j\neq k}^{K}}|\mathbf{h}_{k}\mathbf{w}_{j}|^{2}+WN_{0})}\label{eq:rate:epi}\\
 & \ \ \ {\textstyle \sum_{k=1}^{K}}x_{b,k}z_{k}\leq\bar{C}\label{eq:BH:reform}\\
 & \ \ \ \eqref{eq:soft power},\eqref{eq:power},\eqref{eq:min:connectivity},\eqref{eq:boolean},\eqref{eq:QoS:SoCP},\vspace{-3mm}\label{eq:group-constraints-1}
\end{align}
\end{subequations} where ${\bf z}\triangleq[z_{1},\ldots,z_{K}]^{T}$.
The equivalence between \eqref{Prob:Gen:Problem} and \eqref{Prob:SE:epi}
can be verified as \eqref{eq:rate:epi} holds with equality at the
optimum. Now we can observe that a better objective of \eqref{Prob:SE:epi}
is always achieved if we keep increasing each of $z_{k}$ as long
as it is still in the feasible set of \eqref{Prob:SE:epi}. In addition,
the feasibility of ${\bf z}$ is established depending on ${\bf x}$
as can be seen in \eqref{eq:BH:reform}. Furthermore, constraint \eqref{eq:BH:reform}
is monotone w.r.t. ${\bf z}$ and ${\bf x}$. These three observations
imply that \eqref{Prob:SE:epi} is suitable for a direct application
of the DBRB to find the optimal solutions of ${\bf z}$ and ${\bf x}$.
Before describing the details, we introduce new notations for the
ease of exposition. Let ${\bf s}=[{\bf x}^{T},{\bf z}^{T}]^{T}\in\mathbb{R}_{+}^{N_{x}+N_{z}}$
be the variable vector of interest where $N_{x}=BK$ and $N_{z}=K$
are dimensions of the Boolean and continuous vectors ${\bf x}$ and
${\bf z}$, respectively. Let ${\cal S}$ be the feasible set of problem
\eqref{Prob:SE:epi}, i.e., $\begin{alignedat}{1}{\cal S}=\{{\bf s}\in\mathbb{R}_{+}^{N}\ | & \ \eqref{eq:soft power},\eqref{eq:power},\eqref{eq:min:connectivity},\eqref{eq:boolean},\eqref{eq:QoS:SoCP},\eqref{eq:rate:epi},\eqref{eq:BH:reform}\}\end{alignedat}
$ where $N=N_{x}+N_{z}$ and ${\cal S}$ is normal and finite since
it is upper bounded by the power and backhaul constraints. Additionally,
the feasible set ${\cal S}$ is contained in a box $D=[{\bf a},{\bf b}]\in\mathbb{R}_{+}^{N}$
whose vertices are determined as follows. It is easily seen that
$a_{i}=0$ and $b_{i}=1$ for $i=1,\ldots N_{x}$ since $x_{b,k}\in\{\underline{x}_{b,k},\overline{x}_{b,k}\}=\{0,1\}$.
On the other hand, variable $z_{k}$ is bounded below by $z_{k}\geq\underline{z}_{k}=\log(1+\bar{\gamma}_{0})$.
In addition, we can verify $z_{k}\leq\overline{z}_{k}=\min\{B\bar{C},\log(1+|{\bf h}_{k}{\bf w}_{k}|^{2}/WN_{0})\}\leq\min\{B\bar{C},\log(1+\|{\bf h}_{k}\|_{2}^{2}\|{\bf w}_{k}\|_{2}^{2}/WN_{0})\}\leq\min\{B\bar{C},\log(1+B\bar{P}\|{\bf h}_{k}\|_{2}^{2}/WN_{0})\}$.
That is to say, for $i=N_{x}+1,\ldots,N$, we have $a_{i}=\log(1+\bar{\gamma}_{0})$
and $b_{i}=\min\{B\bar{C},\ \log(1+B\bar{P}\|{\bf h}_{k}\|_{2}^{2}/WN_{0})\}$
where $k=i-N_{x}$. To sum up, the lower and upper vertices of box
$D$ are given by\vspace{-2mm}
\begin{align*}
 & {\bf a}=[{\bf 0}_{N_{x}},\ \log(1+\bar{\gamma}_{0})\times{\bf 1}_{N_{z}}]\\
 & {\bf b}=[{\bf 1}_{N_{x}},\ \left\{ \min\{B\bar{C},\ \log(1+B\bar{P}\|{\bf h}_{k}\|_{2}^{2}/WN_{0})\}\right\} _{k=1}^{N_{z}}],\vspace{-3mm}
\end{align*}
where ${\bf 0}_{N_{x}}$, $\bm{1}_{N_{x}}$ and ${\bf 1}_{N_{z}}$
denote vectors of all zero and one values of the size given in the
subscripts, respectively. \foreignlanguage{english}{}At this line,
problem \eqref{Prob:SE:epi} can be compactly rewritten as \vspace{-2mm}
\begin{equation}
\max\{f({\bf s})\triangleq{\textstyle \sum_{i=N_{x}+1}^{N}}s_{i}\ |\ {\bf s}\in{\cal S}\subset D\}.\vspace{-2mm}\label{Prob:compact form:SE}
\end{equation}
We are now ready to describe the DBRB algorithm to solve \eqref{Prob:compact form:SE}
optimally.  Generally, this method is an iterative procedure consisting
of three basic operations at each iteration:\emph{ branching}, \emph{reduction,
}and \emph{bounding}. More specifically, starting from the box $[{\bf a},{\bf b}]$,
we iteratively divide it into smaller and smaller ones, remove boxes
that do not contain an optimal solution, search over remaining boxes
for a better optimal solution until fulfilling the stopping criterion.
Different to the continuous procedure, to guarantee the exact solution
of the Boolean variable, $N_{x}$ first elements on the cutting plane
of boxes are adjusted to be dropped in the Boolean set during the
branching and reduction operations. The adjustment rule is motivated
by the monotonicity property to ensure not cutting off any feasible
solution \cite{tuy2006discrete}. The algorithm terminates when the
size of boxes containing the optimal solution is small enough. DBRB
algorithm is described in Alg. \ref{Alg. BRB} where details are given
as follows. For notational convenience, we denote by $\vartheta_{n}$,
${\cal R}_{n}$, $f_{U}(V)$ and $f_{L}(V)$ the current best objective
(\emph{cbo}), the set of boxes containing an optimal solution at iteration
$n$, the upper bound and lower bound value of $f({\bf s})$ over
box $V$, respectively.\vspace{-2mm}

\subsubsection*{Branching}

We start iteration $n$ by selecting a box in ${\cal R}_{n}$ and
splitting it into two smaller ones. A candidate box $V_{c}\in{\cal R}_{n}$
for branching is picked up by the improving bound rule \cite{tuy2005monotonic},
i.e., $V_{\text{c}}=\arg\max_{V\in{\cal R}_{n}}\ f_{U}(V)$. The selected
box $V_{\text{c}}=[{\bf p},{\bf q}]$ is then bisected along the longest
edge, i.e., $j=\arg\max_{1\leq i\leq N}(q_{i}-p_{i})$, to create
two new boxes which are of equal size as \vspace{-2mm}
\begin{equation}
\begin{alignedat}{1}V_{\text{c}}^{(1)}=\begin{cases}
[{\bf p},\ {\bf q}-{\bf e}_{j}] & \bm{\text{if }}j\leq N_{x},\\
{}[{\bf p},\ {\bf q}-{\bf e}_{j}(q_{i}-p_{i})/2] & \bm{\text{if }}j>N_{x},
\end{cases}\\
V_{\text{c}}^{(2)}=\begin{cases}
[{\bf p}+{\bf e}_{j},\ {\bf q}] & \bm{\text{if }}j\leq N_{x},\\
{}[{\bf p}+{\bf e}_{j}(q_{i}-p_{i})/2,\ {\bf q}] & \bm{\text{if }}j>N_{x}.
\end{cases}
\end{alignedat}
\label{eq:bisection rule}
\end{equation}
Rule \eqref{eq:bisection rule} ensures that the elements on the cutting
plane corresponding to the Boolean variable are adjusted to be in
the Boolean set. For a resulting box $V_{\text{c}}^{(l)},\ l=\{1,2\}$,
it possibly contains segments which are either infeasible solutions
to \eqref{Prob:compact form:SE} or solutions resulting in a smaller
objective than $\vartheta_{n}$. Thanks to the monotonicity property,
we can remove those portions of no interest by a cutting procedure
referred as reduction operation.\vspace{-2mm}

\subsubsection*{Reduction }

Suppose that the input of this operation is box $\tilde{V}=[{\bf p},{\bf q}]$
and $\tilde{V}$ is assumed to contain an optimal solution. We aim
at reducing the size of the solution set without loss of optimality
by searching for a smaller box $V^{\prime}=[{\bf p}^{\prime},{\bf q}^{\prime}]$,
i.e., $[{\bf p}^{\prime},{\bf q}^{\prime}]\subset[{\bf p},{\bf q}]$
such that an optimal solution must be contained in $V^{\prime}$.
That is, if all vectors belonging to portion $[{\bf p},{\bf p}^{\prime})$
result in a smaller objective value ($f({\bf s})<\vartheta_{n}$)
and/or be outside the feasible set of \eqref{Prob:compact form:SE}
(${\bf s}\in D\backslash{\cal S}$), the portion $[{\bf p},{\bf p}^{\prime})$
must be cut off. On the other hand, we remove the portion $({\bf q}^{\prime},{\bf q}]$
if any vector in the set is infeasible to \eqref{Prob:compact form:SE}.
Mathematically, for each $i=1,\ldots,N$, we can replace ${\bf p}$
by ${\bf p}^{\prime}\geq{\bf p}$ where ${\bf p}^{\prime}={\bf q}-{\textstyle \sum_{i=1}^{N}}\alpha_{i}(q_{i}-p_{i})$
and \vspace{-2mm}
\begin{equation}
\begin{alignedat}{1}\alpha_{i}=\sup\{\alpha\ |0\leq\alpha\leq1, & \ {\bf q}-\alpha(p_{i}-q_{i}){\bf e}_{i}\in D\backslash{\cal S}\\
 & f({\bf q}-\alpha(p_{i}-q_{i}){\bf e}_{i})\geq\vartheta_{n}\}.
\end{alignedat}
\vspace{-2mm}\label{eq:find alpha}
\end{equation}
 Similarly, vertex set ${\bf q}$ is replaced by ${\bf q}^{\prime}\leq{\bf q}$
where ${\bf q}^{\prime}={\bf p}^{\prime}+{\textstyle \sum_{i=1}^{N}}\beta_{i}(q_{i}-p_{i}^{\prime}){\bf e}_{i}$
and\vspace{-2mm}
\begin{equation}
\begin{alignedat}{1}\beta_{i}=\sup\{\beta\ | & 0\leq\beta\leq1,\ {\bf p}^{\prime}+\beta(q_{i}-p_{i}^{\prime}){\bf e}_{i}\in{\cal S}\}\end{alignedat}
\vspace{-2mm}\label{eq:find beta}
\end{equation}
The values of $\alpha_{i}$ and $\beta_{i}$ in \eqref{eq:find alpha}
and \eqref{eq:find beta} can be found easily by the bisection method.
For $i=1,\ldots,N_{x}$, the output of the reduction task should be
nested in the Boolean set, i.e., $p_{i}^{\prime},q_{i}^{\prime}\in\{0,1\}$
since they correspond to the Boolean variable. Thus, by replacing
$q_{i}-p_{i}=1$ into \eqref{eq:find alpha} for $i=1,\ldots,N_{x}$,
we can quickly achieve that $p_{i}^{\prime}=\begin{cases}
1 & \text{if }\ {\bf q}-{\bf e}_{i}\in D\backslash{\cal S}\\
0 & \text{otherwise},
\end{cases}$. If it results in $p_{i}^{\prime}=0$, we then replace $q_{i}-p_{i}^{\prime}=1$
into \eqref{eq:find beta} and achieve $q_{i}^{\prime}=\begin{cases}
1 & \text{if }\ {\bf p}^{\prime}+{\bf e}_{i}\in{\cal S}\\
0 & \text{otherwise}
\end{cases}$.  As have been proved in \cite{tuy2006discrete} that the reduction
procedure above does not drop off any feasible solution of \eqref{Prob:compact form:SE}.
We refer the output of the reduction operation with the input box
$\tilde{V}=[{\bf p},{\bf q}]$ as $\re([{\bf p},{\bf q}])$.
\begin{algorithm}[t]
\caption{The proposed DBRB algorithm }
\label{Alg. BRB}

\begin{algorithmic}[1]

\STATE \textbf{Initialization:} Compute ${\bf a}$, ${\bf b}$ and
apply box reduction to box $[{\bf a},{\bf b}]$. Let $n:=1$, ${\cal R}_{1}=\re([{\bf a},{\bf b}])$
and $\vartheta_{1}=K\log(1+\bar{\gamma}_{0})$

\REPEAT[\textbf{$n:=n+1$}.]

\STATE{\textbf{Branching:} select a box $V_{\text{c}}=[{\bf p},{\bf q}]$
and branch $V_{\text{c}}$ into two smaller ones $V_{\text{c}}^{(1)}$
and $V_{\text{c}}^{(2)}$}

\STATE{\textbf{Reduction}: apply box reduction to each box $V_{\text{c}}^{(l)}$($l=\{1,2\})$
and obtain reduced box $\re(V_{\text{c}}^{(l)})$ }

\STATE{\textbf{Bounding}: For each box $\re(V_{\text{c}}^{(l)})$,
if \eqref{eq:SE:SOCP:bound} is feasible.\label{Bounding}

-Use Alg. \ref{Alg-search} to find a feasible solution, obtain $f_{L}(\re(V_{\text{c}}^{(l)}))$
and update $\vartheta_{n}=\max\{f_{L}(\re(V_{\text{c}}^{(l)})),\vartheta_{n-1}\}$

-Update $f_{U}(\re(V_{\text{c}}^{(l)}))$ and ${\cal R}_{n}={\cal R}_{n-1}\cup\{\re(V_{\text{c}}^{(l)})|f_{U}(\re(V_{\text{c}}^{(l)}))\geq\vartheta_{n}\}$}

\UNTIL{Convergence}

\end{algorithmic}
\end{algorithm}
\vspace{-3mm}

\subsubsection*{Bounding}

\vspace{-0.5mm}The bounding operation aims at updating the upper
and lower bounds of the resulting boxes from the reduction operator,
thereby removing boxes of no interest whose upper bound is smaller
than the \emph{cbo}. The upper and lower bounds of box $\re([{\bf p},{\bf q}])=[{\bf p}^{\prime},{\bf q}^{\prime}]$
can be simply computed as $f({\bf q}^{\prime})$ and $f({\bf p}^{\prime})$,
respectively, due to the monotonic increase of the objective. However,
this often results in slow convergence rate. Instead, we now consider
a better bound computation for box $[{\bf p}^{\prime},{\bf q}^{\prime}]$
as follows. Recall that problem \eqref{Prob:SE:epi} is NP hard
due to the Boolean variable ${\bf x}$ and nonconvex constraints \eqref{eq:rate:epi},
\eqref{eq:BH:reform}. Nevertheless, we can compute the upper bound
of $f({\bf s})$ by a convex relaxation of \eqref{Prob:SE:epi}. That
is, we replace the left side of \eqref{eq:BH:reform} by its convex
envelope, i.e., ${\textstyle \sum_{k=1}^{K}}x_{b,k}z_{k}\geq\phi_{b}(x_{b,k},z_{k})$
where $\phi_{b}(x_{b,k},z_{k})\triangleq\max\{{\textstyle \sum_{k=1}^{K}}(\underline{z}_{k}x_{b,k}+\underline{x}_{b,k}z_{k}-\underline{x}_{b,k}\underline{z}_{k});{\textstyle \sum_{k=1}^{K}}(\overline{z}_{b}x_{b,k}+\overline{x}_{b,k}z_{k}-\overline{x}_{b,k}\overline{z}_{k})\}$
for $z_{k}\in[\underline{z}_{k},\overline{z}_{k}]$ and $x_{b,k}\in[\underline{x}_{b,k},\overline{x}_{b,k}]$
\cite{convex-concave-envelope}. Note that $(\underline{x}_{b,k},\underline{z}_{k})$
and $(\overline{x}_{b,k},\overline{z}_{k})$ correspond to elements
in ${\bf p}^{\prime}$ and ${\bf q}^{\prime}$, respectively. In addition,
the right side of \eqref{eq:rate:epi} can be replaced by its lower
bound as
\begin{equation}
\mathbf{h}_{k}\mathbf{w}_{k}\geq\sqrt{(e^{\underline{z}_{k}}-1)({\textstyle \sum_{j\neq k}^{K}}|\mathbf{h}_{k}\mathbf{w}_{j}|^{2}+WN_{0})}.\label{eq:rate:bound}
\end{equation}
Then, by treating ${\bf x}$ as a continuous variable vector, the
upper bound of $f({\bf s})$ is computed by solving the following
SOCP problem \begin{subequations}\label{eq:SE:SOCP:bound}\vspace{-2mm}
\begin{align}
\max_{{\bf w},{\bf x},{\bf u},{\bf z}} & \ \ \ {\textstyle \sum}_{k=1}^{K}z_{k}\\
\text{subject to} & \ \ \ \eqref{eq:power},\eqref{eq:min:connectivity},\eqref{eq:QoS:SoCP},\eqref{eq:rate:bound},\eqref{eq:BH:reform}\\
 & \ \ \ \|{\bf w}_{b,k}^{T}\ \ (x_{b,k}-u_{b,k})/2\|_{2}\leq(x_{b,k}+u_{b,k})/2,\label{eq:soft power:SOCP}\\
 & \ \ \ \vartheta_{n}\leq{\textstyle \sum}_{k=1}^{K}z_{k}\leq{\textstyle \sum}_{k=1}^{K}\overline{z}_{k},\\
 & \ \ \ \phi_{b}(x_{b,k},z_{k})\leq\bar{C},\vspace{-2mm}
\end{align}
\end{subequations}where \eqref{eq:soft power:SOCP} is the SOC representation
of \eqref{eq:soft power} when $x_{b,k}\in[0,1]$. Let $f({\bf s}^{\ast})=\sum_{k=1}^{K}z_{k}^{\ast}$
be the optimal objective where $(.)^{*}$ denotes the solution of
\eqref{eq:SE:SOCP:bound}. Although ${\bf s}^{\ast}$ may not be feasible
to \eqref{Prob:SE:epi} as discrete constraint \eqref{eq:boolean}
is neglected, we can still achieve that $f({\bf s}^{\ast})\leq f({\bf q}^{\prime})$
\cite{convex-concave-envelope}. In addition, we can also compute
a better lower bound if a feasible solution to \eqref{Prob:SE:epi}
in box $[{\bf p}^{\prime},{\bf q}^{\prime}]$ (denote as $\hat{{\bf s}}$)
is determined, i.e., $f(\hat{{\bf s}})\geq f({\bf p}^{\prime})$.
We can establish $\hat{{\bf s}}$ by some insights gained from the
optimal solution to \eqref{eq:SE:SOCP:bound}. Particularly, we can
see that the smallest elements of soft power level ${\bf u}^{\ast}$
imply the less contribution of corresponding transmission links in
satisfying \eqref{eq:rate:bound}. To this point, a feasible selection
vector ${\bf x}$ may be determined by turning off those transmission
links (i.e., force $x_{b,k}^{\ast}=0$ if $u_{b,k}^{\ast}$ is small
enough and set the remaining ones $x_{b,k}^{\ast}=$1). On the other
hand, given a pre-determined selection vector $\tilde{{\bf x}}\in\{0,1\}^{N_{x}}$,
it is said to be feasible if ${\textstyle \sum_{b=1}^{B}}\tilde{x}_{b,k}\geq1$
and there exists ${\bf w}$ and ${\bf u}$ which satisfying the following
constraints \begin{subequations}
\begin{gather}
\|{\bf w}_{b,k}^{T}\ (\tilde{x}_{b,k}-u_{b,k})/2\|_{2}\leq(\tilde{x}_{b,k}+u_{b,k})/2,\ \eqref{eq:power},\ \eqref{eq:rate:bound},\label{sub:SE:checkfeasible}\\
{\textstyle \sum_{k=1}^{K}}\tilde{x}_{b,k}R({\bf w})\leq\bar{C}.\vspace{-3mm}\label{eq:rechek:BH}
\end{gather}
\end{subequations} With these observations, we can derive a binary-search-based
approach to find an optimal solution in box $[{\bf p}^{\prime},{\bf q}^{\prime}]$.
The central idea is to iteratively pick vector $\tilde{{\bf x}}$
based on solution ${\bf u}^{\ast}$ of \eqref{eq:SE:SOCP:bound},
and verify its feasibility by solving the problem $\{\bm{\text{find}}_{{\bf w},{\bf u}}\ \text{s.t.}\ \eqref{sub:SE:checkfeasible}\}$
and (if feasible) checking the obtained solution with \eqref{eq:rechek:BH}.
The algorithm outputs a solution that yields the best objective among
all validated feasible solutions in box $[{\bf p}^{\prime},{\bf q}^{\prime}]$.
We use this solution to update the lower bound $f_{L}([{\bf p}^{\prime},{\bf q}^{\prime}])$.
Details of the searching method is described in Alg.\ \ref{Alg-search}
and used at step \ref{Bounding} of the DBRB algorithm (Alg.\ \ref{Alg. BRB}).
The convergence property of Alg.\ \ref{Alg. BRB} is followed by
the one in \cite{tuy2006discrete} and numerically shown in next section.
 \vspace{-2mm}

\section{Numerical Results}

\vspace{-1mm}
\begin{algorithm}[b]
\caption{The binary search algorithm}
\label{Alg-search}

\begin{algorithmic}[1]

\STATE \textbf{Initialization:}\label{Alg:search:initialization}
Let ${\bf u}^{\ast}$ be the solution of \eqref{eq:SE:SOCP:bound}
. Set $L_{\min}=K$ and $L_{\max}=N_{x}$. Set $f^{\text{opt}}=\sum_{k=1}^{K}\underline{z}_{k}$

\WHILE{$L_{\min}<L_{\max}$}

\STATE{Set $L=\left\lfloor (L_{\max}+L_{\min})/2\right\rfloor $}\label{choosing L}

\STATE{Let $\lambda_{L}$ be value of the $L$-th largest element
of ${\bf u}^{\ast}$.}

\STATE{Set $\tilde{x}_{b,k}=1$ if $u_{b,k}^{\ast}\geq\lambda_{L}$
and $\tilde{x}_{b,k}=0$ otherwise.}\label{pick x}

\IF{Solving $\{\bm{\text{find}}_{{\bf w},{\bf u}}\ \text{s.t.}\ \eqref{sub:SE:checkfeasible}\}$
is feasible}

\STATE{Obtain solution $\hat{{\bf w}}$ and calculate achieved $R_{k}(\hat{{\bf w}})$
}

\IF{$\sum_{k=1}^{K}\tilde{x}_{b,k}R_{k}(\hat{{\bf w}})\leq\bar{C},\ \forall b,k$
}

\STATE{Set $f^{\text{opt}}=\max\{f^{\text{opt}},\sum_{k=1}^{K}R_{k}(\hat{{\bf w}})\}$}\label{update opt}

\ENDIF

\ELSE\STATE{Set $L_{\max}=L-1$ and return to step \ref{choosing L}
}

\ENDIF

\STATE{Set $L_{\min}:=L+1$}\label{increase L}

\ENDWHILE\label{endwhile}

\end{algorithmic}
\end{algorithm}
\begin{figure}[h]
\centering{}\includegraphics[width=0.8\columnwidth]{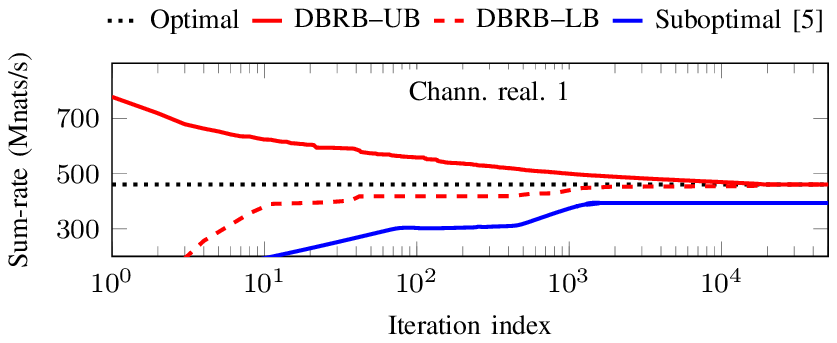}\smallskip{}
\includegraphics[width=0.8\columnwidth]{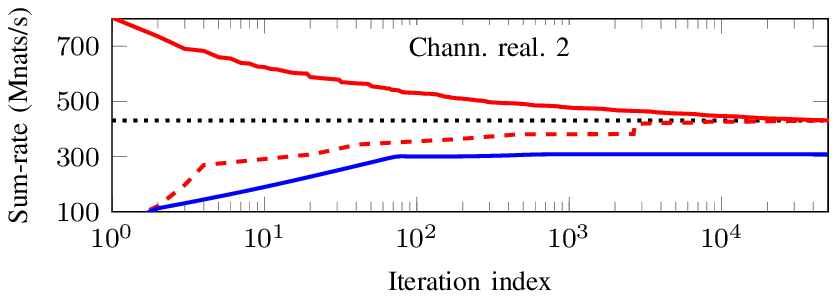}\vspace{-2mm}\caption{Convergence behavior of the DBRB and comparison to the suboptimal
one in \cite{Dai2014} for two channel realizations with $\bar{C}=200$
Mnats/s.}
\label{fig. 1}
\end{figure}
\begin{figure}[t]
\centering{}\includegraphics[width=0.81\columnwidth]{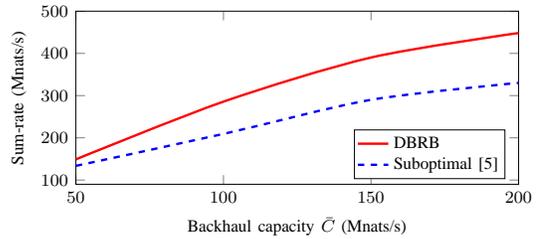}\vspace{-2mm}\caption{Average sum rate versus the backhaul capacity $\bar{C}$. }
\label{fig. 2}\vspace{-4mm}
\end{figure}
We numerically evaluate the proposed DBRB method. We consider a network
consisting of $B=3$ BSs equipped by $M=4$ antennas and $K=6$ single-antenna
users randomly placed in the coverage area of all BSs. The inter-site
distance between two BSs is $d=1$\,km. The pathloss model is given
by $\text{PL(dB)}=128.1+37.6\log_{10}(d)$ and the standard deviation
of the log normal shadowing is 8. The transmit power budget is $\bar{P}=46$
dBm, the noise power density is $N_{0}=-174$ dBm/Hz and the system
bandwidth is $W=10$ MHz. We also set the per-user specified SINR
$\bar{\gamma}_{0}=0$ dB. For comparison purposes, we use Alg.\,\ref{Alg. BRB}
as a benchmark to the suboptimal method studied in \cite{Dai2014}
which measures the same backhaul metric\,as\,this paper.

Fig.\,\ref{fig. 1} shows examples of the convergence behavior of
the DBRB algorithm, i.e., the convergence is declared when the gap
between upper bound (UB) and lower bound (LB) is small enough. As
can be seen, this gap is reduced rapidly during first few iterations
since a large number of infeasible portions are cut off. Alg.\,\ref{Alg. BRB}
converges after a finite number of iterations. Another interesting
observation is that the performance of the solution in \cite{Dai2014}
is quite far from the optimal one. This can also be seen by Fig.\,\ref{fig. 2}
where we illustrate the average sum rate versus the backhaul capacity. Fig.\,\ref{fig. 2} demonstrates that
the suboptimal method only attains 70\% to 90\% of the optimal performance.
Thus, there is a need of a more efficient low-complexity scheme. \vspace{-2mm}

\section{Conclusion}

\vspace{-1mm}This paper has considered the problem of joint beamforming
and BS-user link selection to maximize the sum rate in a downlink
CoMP transmission with limited backhaul capacity links. We have derived
an optimization framework that solves the design problem to global
optimality by customizing a DBRB algorithm. We have also numerically
shown the finite convergence of the proposed method. The proposed
optimal solution serves as a benchmark and can generate design guidelines
for the suboptimal solutions. The DBRB framework can be extended for
similar problems involving different backhaul usage measures.


\end{document}